\documentclass[aps,pra,twocolumn,superscriptaddress,floatfix]{revtex4}
\usepackage{amsmath}
\usepackage{amsfonts}
\usepackage{amssymb}
\usepackage{graphicx}
\usepackage{cancel}
\usepackage[switch,columnwise]{lineno}
\usepackage{titlesec}
\usepackage{xcolor}
\usepackage{hyperref}

\makeatletter
\renewcommand{\section}{\@startsection{section}{1}{0mm}
{-\baselineskip}{0.5\baselineskip}{\bf\leftline}}

\renewcommand{\subsection}{\@startsection{section}{1}{0mm}
{-\baselineskip}{0.5\baselineskip}{\bf\leftline}}
\makeatother

\begin{document}

\title{Retrodiction beyond the Heisenberg uncertainty relation}

\author{Han Bao}%
\affiliation{State Key Laboratory of Quantum Optics and Quantum Optics Devices, Institute of Laser Spectroscopy, Collaborative research center on Quantum optics and extreme optics, Shanxi University, Taiyuan, Shanxi 030006, China}%
\affiliation{Department of Physics, State Key Laboratory of Surface Physics and Key Laboratory of Micro
and Nano Photonic Structures (Ministry of Education), Fudan University, Shanghai 200433, China}%
\author{Shenchao Jin}%
\affiliation{Department of Physics, State Key Laboratory of Surface Physics and Key Laboratory of Micro
and Nano Photonic Structures (Ministry of Education), Fudan University, Shanghai 200433, China}%
\author{Junlei Duan}%
\affiliation{Department of Physics, State Key Laboratory of Surface Physics and Key Laboratory of Micro
and Nano Photonic Structures (Ministry of Education), Fudan University, Shanghai 200433, China}%
\author{Suotang Jia}%
\affiliation{State Key Laboratory of Quantum Optics and Quantum Optics Devices, Institute of Laser Spectroscopy, Collaborative research center on Quantum optics and extreme optics, Shanxi University, Taiyuan, Shanxi 030006, China}%
\author{Klaus M\o lmer}%
\email{moelmer@phys.au.dk}
\affiliation{Department of Physics and Astronomy, Aarhus University, Ny Munkegada 120, DK-8000 Aarhus C. Denmark}%
\author{Heng Shen}%
\email{heng.shen@physics.ox.ac.uk}
\affiliation{State Key Laboratory of Quantum Optics and Quantum Optics Devices, Institute of Opto-electronics,  Collaborative research center on Quantum optics and extreme optics, Shanxi University, Taiyuan, Shanxi 030006, China}%
\affiliation{Clarendon Laboratory, University of Oxford, Parks Road, Oxford, OX1 3PU, UK}%
\author{Yanhong Xiao}%
\email{yxiao@fudan.edu.cn}
\affiliation{State Key Laboratory of Quantum Optics and Quantum Optics Devices, Institute of Laser Spectroscopy, Collaborative research center on Quantum optics and extreme optics, Shanxi University, Taiyuan, Shanxi 030006, China}%
\affiliation{Department of Physics, State Key Laboratory of Surface Physics and Key Laboratory of Micro
and Nano Photonic Structures (Ministry of Education), Fudan University, Shanghai 200433, China}%
%

\begin{abstract}
In quantum mechanics, the Heisenberg uncertainty relation presents an ultimate limit to the precision by which one can predict the outcome of position and momentum measurements on a particle. Heisenberg explicitly stated this relation for the prediction of ``hypothetical future measurements", and it does not describe the situation where knowledge is available about the system both earlier and later than the time of the measurement. We study what happens under such circumstances with an atomic ensemble containing $10^{11}$ $^{87}\text{Rb}$ atoms, initiated nearly in the ground state in presence of a magnetic field. The collective spin observables of the atoms are then well described by canonical position and momentum observables, $\hat{x}_A$ and $\hat{p}_A$ that satisfy $[\hat{x}_A,\hat{p}_A]=i\hbar$. Quantum non-demolition measurements of $\hat{p}_A$ before and of $\hat{x}_A$  after time $t$ allow precise estimates of both observables at time $t$. The capability of assigning precise values to multiple observables and to observe their variation during physical processes may have implications in quantum state estimation and sensing.
\end{abstract}
\maketitle

\begin{figure*}
\centering
\includegraphics[width=0.8\textwidth]{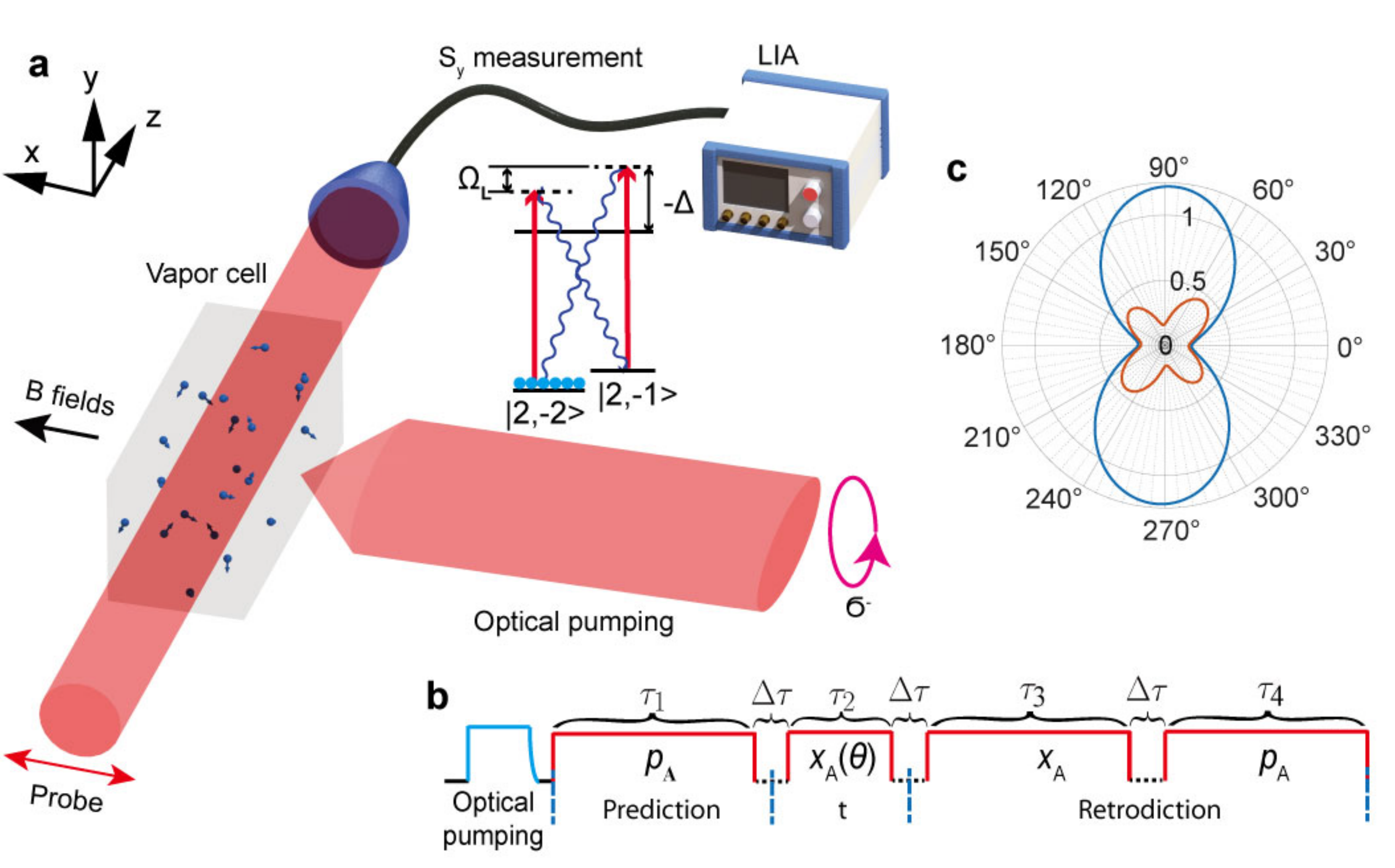}
\caption{\label{Fig:setup} \textbf{Schematics for retrodiction beyond the Heisenberg uncertainty relation}.
\textbf{a.} Experiment schematics. A paraffin coated $20\text{mm}\times7\text{mm}\times7\text{mm}$ rectangular vapor cell at $53^{\circ}C$ resides inside a 4-layer magnetic shielding. A coherent spin state (CSS) is created along the quantization axis \emph{x} by optical pumping, with a pump laser tuned to the Rb D1 transition $|5S_{1/2},F=2\rightarrow5P_{1/2},F'=2$ and a repump laser beam stabilized to the Rb D2 transition $5S_{1/2},F=1\rightarrow5P_{3/2},F'=2$, sharing the same circular polarization. A magnetic field of $0.71~\text{G}$ induces a ground-state Zeeman splitting of about $\Omega_L=2\pi\times500~\text{kHz}$ and stabilizes the collective spin component along the \emph{x} direction. A linearly polarized off-resonance D2 laser beam, propagating in the \emph{z} direction, probes the quantum fluctuations of the spin. The Stokes component $S_{y}$ that characterizes the linear polarization in the $\pm 45^{\circ}$ basis is measured using a balanced polarimetry scheme. A Lock-In amplifier (LIA) extracts the signal at the Lamor frequency. \textbf{b.} Pulse sequence. The pump lasers prepare the atoms in the CSS. The pump lasers are turned off and the probe laser is turned on to interact with the atoms. The probe pulse sequence is divided in four parts, measuring the quadratures indicated in the figure. \textbf{c.} Theoretical polar plot of the variance of the atomic oscillator quadratures conditioned on prior (blue curve) and on prior and posterior measurements (red curve). The plot assumes ideal experiments without decoherence and decay, probing strengths $\kappa_1^2= 1.7$, $\kappa_3^2= 3.3$ and $\kappa_4^2=2.2$ and a duty factor of $14\%$, for details see Supp.Mat. The radial distance of $0.5$ in the polar plot represents the SQL, and the retrodicted variances of the horizontal and vertical $\hat{p}_A$ and $\hat{x}_A$ quadratures are both reduced in violation of the HUR.}
\end{figure*}

Heisenberg's uncertainty relation (HUR) \cite{HUR} is one of the pillars of quantum mechanics and it sets the limit of how precisely one can predict the outcome of the measurements of two non-commuting observables. While the relation itself is a simple consequence of Born's rule and the operator character of physical observables, it has spurred both foundational discussions of the interpretation of quantum theory and efforts to identify and surpass quantum limits for practical high precision measurements. The HUR deals with the ability to predict the outcomes of measurements of either one of two observables in different experiments with the same quantum state.  As a related concept in quantum metrology, the so-called standard quantum limit (SQL) denotes the measurement precision achievable with conventional resources, such as coherent states of light and product states of many particles. Poissonian counting statistics and field amplitude measurements with equal size errors on all quadratures are examples of the SQL, while number states and squeezed states permit precision measurements of a single, relevant observable below the SQL \cite{Lloyd}.

Beyond the conventional HUR scenario concerned with measurements of only one observable in each experiment, more complex scenarios have been pursued, where more observables are, simultaneously or sequentially, measured in the same experiment. Measurements generally disturb a quantum system \cite{Ozawa,BuschWerner} and several approaches have been proposed to mitigate effects of this disturbance in high precision measurements of one or more observables over time. These include entanglement based ``negative mass" or ``quantum-mechanics-free" subsystem measurements \cite{Polzik1,HammererPRL,PolzikAnn,hybrid,Tsang} and quantum dense metrology based on an Einstein-Podolsky-Rosen entangled two-mode system \cite{Schnabel}.
The present work is not aimed at such scenarios but serves to illustrate how the HUR does not account for our ability to retrodict the outcome of a measurement on a system at a past time \emph{t}, if we have access to the system after the measurement \cite{Vaidman}. Consider for example the particularly simple situation where prior to the measurement at time t the system was prepared in an eigenstate $|a_m\rangle$ of an observable $\hat{A}$, while a projective measurement of another observable $\hat{B}$ is applied right after \emph{t}, yielding the eigenvalue $b_n$. Clearly, under this circumstance, the outcome of a hypothetical measurement of $\hat{A}$ ($\textit{or}~\hat{B}$) at time t has to be $a_m$ ($\textit{or}~b_n$ in order to be consistent with the subsequent projective measurement) with certainty, even when $\hat{A}$ and $\hat{B}$ do not commute. Separate formalism has been developed to describe the sometimes paradoxical situations occurring in weak and strong measurement scenarios with prior and posterior measurement information \cite{Vaidman,WeakMeasurement,Aharonov}.

Recently, quantum trajectory theory where the density matrix $\rho(t)$ is conditioned on the dynamics and probing of the system until time \emph{t}, was supplemented by an effect matrix $E(t)$ conditioned on the dynamics and probing of the system in a subsequent time interval $[t,T]$ \cite{KlausPRL,KlausPRA}. At the final time $T$, the matrices $\rho(t)$ and $E(t)$ together incorporate all our information about the ``past quantum state", i.e., they yield the probability for the outcome of any general measurement which could have been performed in the laboratory at the earlier time \emph{t}. Analogous to the forward-backward formalism of hidden Markov models and the similar smoothing procedures in Kalman filtering theory, the past quantum state formalism provides an improved estimate of the system dynamics and allows better estimation of physical parameters and of time dependent perturbations on the system. This has been demonstrated by the observed conditional dynamics of the photon number evolution in a cavity \cite{HarochePRA}, the excitation and emission dynamics of a superconducting qubit \cite{Murch} and the motional state of a mechanical oscillator \cite{Albert}. For an alternative application of prior and posterior measurements, see also the definition of quantum smoothing in \cite{Guevara,Laverich}. However, experimental studies of such schemes have so far only addressed retrodiction of a single observable, and hence the prospects of ``violating" the HUR have not yet been demonstrated.

Here, we demonstrate theoretically and experimentally that the past quantum state formalism ``violates" the HUR for a spin oscillator. It is shown that quantum non-demolition (QND) measurements in the earlier and later time intervals $[0,t_-]$ and $[t_+,T]$ provide better estimates of the outcome of measurements at time $t$ of either of the observables than could be inferred from the HUR. We present theory that shows the relation between optical QND measurements results and the inferred spin oscillator variance.

Our experiment addresses the collective atomic spin of an ensemble of $10^{11}$ $^{87}$Rb atoms contained in a macroscopic vapor cell. Previously, we used a similar setup to demonstrate measurement based spin squeezing for a single observable $\hat{p}_A$ and showed that subsequent probing of the same observable improved the ability to guess the outcome of measurements of that quantity further below the SQL~\cite{Bao}. The improved estimation of the expected experimental outcome was then exploited to demonstrate RF magnetic field sensitivity better than the SQL ~\cite{Bao}. In the experiment reported in the present article, we develop a four-pulse measurement of different observables, that allows retrodiction of spin quadratures along any direction, and we show that the ability to guess the outcome of a past position and momentum measurement is not generally limited by any HUR. We can, indeed, infer the outcomes of measurements of both the $\hat{x}_A$ and $\hat{p}_A$ observables with errors below the SQL. This protocol holds potential for estimation of perturbations causing displacements along any directions in phase space, without change of the preparation and post-selection steps. In future application, these may be used in magnetometers equally capable of measuring the amplitude and phase of an RF magnetic field below the SQL.

\noindent
\textbf{Atom-light Faraday interaction.}
Consider the collective atomic spin $\hat{J}_i=\sum_{k=1}^{N_{at}}\hat{j}_{i}^k$, with $i=x, y, z$, given by the sum of the total angular momenta $\hat{j}_{i}^k$ of individual atoms. The macroscopic spin orientation $J_x$ is along the applied bias magnetic field B, and the perpendicular collective spin components $\hat{J}_{y,z}$ oscillate in the lab frame at the Larmor frequency $\Omega_L$.  We denote $\begin{pmatrix}
\hat{J}_{y0}\\
\hat{J}_{z0}
\end{pmatrix}=\begin{pmatrix}
 \cos\Omega_L t&\sin\Omega_L t \\
 -\sin\Omega_L t& \cos\Omega_L t
\end{pmatrix}
\begin{pmatrix}
\hat{J}_{y}\\
\hat{J}_{z}
\end{pmatrix}$ as the spin in the rotating frame. Assuming a highly oriented spin state, the Holstein-Primakoff transformation maps the perpendicular spin operators in the rotating frame to the oscillator quadrature operators $\hat{x}_A=\hat{J}_{y0}/\sqrt{\left | \left \langle J_x \right \rangle \right |}$ and $\hat{p}_A=\hat{J}_{z0}/\sqrt{\left | \left \langle J_x \right \rangle \right |}$. The spin commutator $\left [ \hat{J}_{y0},\hat{J}_{z0} \right ]=iJ_{x} (\hbar=1)$, leads to the HUR, $\Delta\hat{x}_A\cdot \Delta \hat{p}_A \geq 1/2$. The ground state of the harmonic oscillator corresponds to all atoms being in the $|5S_{1/2},F=2,m_F=-2\rangle$ state, forming the Coherent Spin States (CSS) characterized
by $\text{Var}(\hat{J}_{y0})=\text{Var}(\hat{J}_{z0})=J_x/2=N_{at}F/2$ and $\Delta \hat{x}_A \cdot \Delta \hat{p}_A = 1/2$. The first excited state of the oscillator corresponds to a symmetric superposition state of the ensemble with one atom in the state $|5S_{1/2},F=2, m_F=-1\rangle$ \cite{GH,RMP}.

In the limit of large probe detuning with respect to the atomic excited-state hyperfine level \cite{GH,RMP}, the spin observable $\hat{J}_z$ is coupled to the Stokes operators $\hat{S}_z$ of an optical probe pulse with $N_{ph}$ photons and duration $\tau$ via the far off-resonance Faraday interaction $\hat{H}_{int}=(\sqrt{2}\kappa/\sqrt{N_{ph}N_{at}})\hat{J}_{z}\hat{S}_{z}=\frac{\kappa}{\tau}\hat{p}_A\hat{p}_L$, permitting the QND measurement of $\hat{J}_z$ \cite{Braginsky,Thorne,GH,Braginsky3,Schwab}. Here canonical operators of light are defined as $\hat{x}_L=\hat{S}_{y}/\sqrt{\left | \left \langle S_x \right \rangle \right |}$ and $\hat{p}_L=\hat{S}_{z}/\sqrt{\left | \left \langle S_x \right \rangle \right |}$. The coupling constant $\kappa^2\propto N_{ph}N_{at}$ characterizes the strength of the atom-light interaction.

\noindent
\textbf{Conditioned dynamics and the past quantum state.}
The unitary evolution operator of the QND interaction can be written as
\begin{equation}
\hat{U}=e^{-i\hat{H}_{int}\tau}=e^{-i\kappa\hat{p}_A\otimes \hat{p}_L}.
\end{equation}

As explained below, the QND measurement of $\hat{p}_A$ can be generalized to that of a spin component $\hat{x}_A(\theta)=\hat{p}_A\cos\theta+\hat{x}_A\sin\theta$ with an arbitrary direction $\theta$ in the oscillator phase space. After the interaction, the field quadrature $\hat{x}_L$ is measured, which amounts to an indirect, noisy measurement of $\hat{x}_A(\theta)$. Thus, if the field outcome is $m$, the atomic state is transformed by the operator,
\begin{equation}
\begin{aligned}\label{measure_operator}
\hat{\Omega}_{m}=\int \psi_{\hat{x}_L}(m-\kappa a)| a,\theta \rangle\langle a,\theta|\mathrm{d}a
\end{aligned}
\end{equation}
where $| a,\theta \rangle$ is the eigenstate of $\hat{x}_A(\theta)$ with eigenvalue $a$, while
$\psi_{\hat{x}_L}(m)=\frac{1}{\pi^{1/4}}\exp(-\frac{m^2}{2})$ characterizes the quadrature distribution of the input coherent state of the probe laser beam. The operators $\{\hat{\Omega}_{m}\}$ specify a positive operator valued measurement (POVM) \cite{Wiseman-book}).
For infinite $\kappa$, $\hat{\Omega}_{m}$ converges to a projective measurement $\hat{\Omega}_{a}=| a,\theta \rangle\langle a,\theta|$, which projects the atomic state on the eigenstate $| a,\theta \rangle$ with $a=m/\kappa$.

Conditioned upon the output value $m_1$ of the first measurement, the atomic oscillator is described by the unnormalized density matrix $\rho=\Omega_{m_1}\rho_0\Omega_{m_1}^{\dagger}$, with the probability distribution $\operatorname{Pr}(a|m_1)\propto \operatorname{Tr}(\hat{\Omega}_a\rho\hat{\Omega}^{\dagger}_a)= \langle a,\theta|\rho|a,\theta\rangle$ for a subsequent projective measurement of the atomic observable $\hat{x}_A(\theta)$. The distribution is Gaussian and we denote its expectation value and variance by $\mu_{\rho}(\theta)$ and $\sigma^2_{\rho}(\theta)$. The conditional variance does not depend on the outcome of the first measurement.

In the experiment, however, we are restricted to optical measurements employing the finite Faraday interaction described by the POVM $\hat{\Omega}_{m_2}=\int \psi_{\hat{x}_L}(m_2-\kappa_2 a)| a,\theta \rangle\langle a,\theta|\mathrm{d}a$, and the corresponding distribution $\operatorname{Pr}(m_2|m_1)\propto \operatorname{Tr}(\hat{\Omega}_{m_2}\rho\hat{\Omega}^{\dagger}_{m_2})$. This interaction imprints the atomic observable onto the light observable which acquires the expectation value $\langle m_2\rangle=\kappa_2\mu_{\rho}(\theta)$, while shot noise fluctuations contribute to the variance of the optical measurement, $\operatorname{Var}(m_2|m_1)=\kappa_2^2\sigma^2_{\rho}(\theta)+ \frac{1}{2}$ (see Supplementary Materials). It is  customary to regard the optical measurement as a noisy measurement of the atomic observable and thus infer its variance by the relation, $\sigma^2_{\rho}(\theta) = (\operatorname{Var}(m_2|m_1)- \frac{1}{2})/\kappa_2^2$.

So far, our discussion was concerned with the usual application of the conditioned quantum state to determine the uncertainty of the outcome of projective and general measurements. Now, we turn to the case where such outcomes are retrodicted by the combination of a prior measurement of $\hat{p}_A$ with measurement strength $\kappa_1$ and posterior measurements of $\hat{x}_A$ and $\hat{p}_A$ with measurement strengths $\kappa_3$ and $\kappa_4$, respectively. The reason we implement the fourth measurement is explained in Supplementary Materials. If we consider a projective second measurement, represented by $\hat{\Omega}_a$, the joint probability of all four measurement outcomes is $\operatorname{Tr}\left(\hat{\Omega}_{m_4}\hat{\Omega}_{m_3}\hat{\Omega}_{a}\hat{\Omega}_{m_1} \rho_0 \hat{\Omega}_{m_1}^{\dagger}\hat{\Omega}_{a}^{\dagger}\hat{\Omega}_{m_3}^{\dagger}\hat{\Omega}_{m_4}^{\dagger}\right)$.
Fixing the arguments of this expression by the known values of $m_1,\ m_3$ and $m_4$, this yields the conditional probability $\text{Pr}(a|m_1,m_3,m_4)\propto \operatorname{Tr}\left(\hat{\Omega}_{a} \rho \hat{\Omega}_{a}^{\dagger}E\right)$, where the effect matrix $E$ is defined as $E=\hat{\Omega}_{m_3}^{\dagger}\hat{\Omega}_{m_4}^{\dagger}\hat{\Omega}_{m_4}\hat{\Omega}_{m_3}$. It follows that $\text{Pr}(a|m_1,m_3,m_4)\propto \langle a,\theta|\rho|a,\theta\rangle \langle a,\theta|E|a,\theta\rangle$, and it is easy to show that by our assumptions,  $\langle a,\theta |E|a,\theta \rangle$ is a Gaussian function. We denote the centroid and variance of this distribution function by $\mu_E(\theta)$ and $\sigma_E^2(\theta)$ (see Supplementary Materials).

We thus assign the probability distribution $\text{Pr}(a|m_1,m_3,m_4)$ to the  outcome of a past projective atomic measurement, and we readily evaluate its expectation value $\mu_{\rho E}(\theta)=\frac{\mu_\rho(\theta)\sigma^2_E(\theta)+\mu_E(\theta)\sigma^2_\rho(\theta)}{\sigma^2_\rho(\theta)+\sigma^2_E(\theta)}$ and variance $\operatorname{Var}(a|m_1,m_3,m_4)=\sigma_{\rho E}^2(\theta)=\frac{1}{1/\sigma_{\rho}^2(\theta)+1/\sigma_E^2(\theta)}$. Polar plots for $\sigma_\rho^2(\theta)$ and $\sigma_{\rho E}^2(\theta)$ based on analytical expressions in Supplementary Materials are shown as blue and red curves in Fig. \ref{Fig:setup} (c).
As $\sigma_\rho^2(\theta=0)$ and $\sigma_{E}^2(\theta=\pi/2)$ may be independently reduced well below $1/2$, $\sigma_{\rho E}^2(\theta)$ may expose squeezing of both $\hat{p}_A$ and $\hat{x}_A$.

While the theory thus shows that the HUR does not apply for retrodiction of projective measurements, we recall that our experiments are based on optical probing, and for a comparison between theory and experiment, we must address the predictions for the POVM $\hat{\Omega}_{m_2}$, conditioned on $m_1$, $m_3$ and $m_4$. They read (See Supplementary Materials),
\begin{equation}\label{offdiag}
\begin{aligned}
&\text{Pr}(m_2 |m_1,m_3,m_4)\\
&\propto\int\int \psi_{\hat{x}_L}(m_2-\kappa_2 a)\psi_{\hat{x}_L}(m_2-\kappa_2 a')\\
&~~~~~~~~~~\cdot\langle a,\theta |\rho| a',\theta \rangle\langle a',\theta |E| a,\theta \rangle\mathrm{d}a\mathrm{d}a'.
\end{aligned}
\end{equation}
and we can show (See Supplementary Materials) that for the special cases of $\theta=0,\frac{\pi}{2}$, we recover the same simple relation between the variances of the atomic and the optical measurements, as we found for the prediction based on the density matrix $\rho$,
\begin{equation} \label{scaling}
\operatorname{Var}(m_2|m_1,m_3,m_4)=\kappa_2^2\sigma_{\rho E}^2(\theta)+ \frac{1}{2},\ (\theta=0,\frac{\pi}{2}).
\end{equation}
The reduced fluctuations of the optical measurements around their retrodicted mean value thus constitute a test of the past quantum state  theory, and if the uncertainty product $\Delta \hat{x}_A\cdot  \Delta \hat{p}_A$ inferred from the optical measurements and Eq.\eqref{scaling} is less than $1/2$, it may be taken as a demonstration of the violation of the HUR.

\begin{figure*}
\centering
\includegraphics[width=0.9\textwidth]{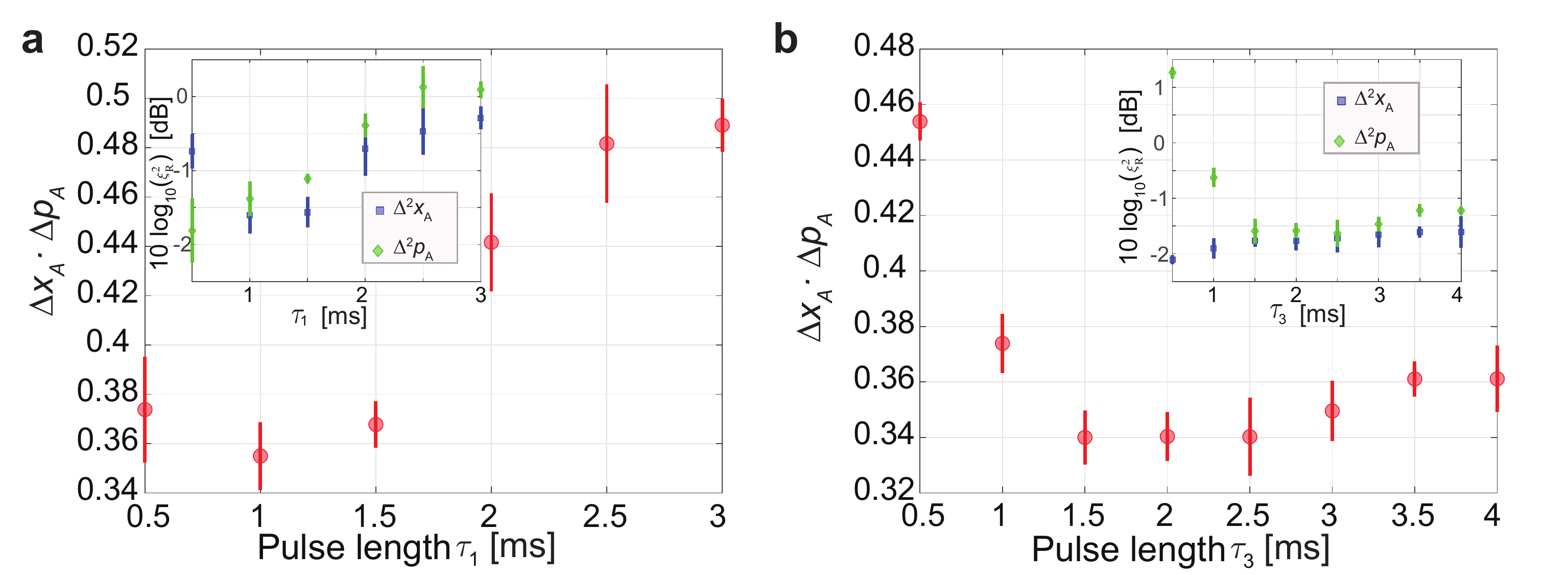}
\caption{\label{Fig:xp} \textbf{Value of $\Delta\hat{x}_A\cdot\Delta\hat{p}_A$ as a function of (a) $\tau_1$ and (b) $\tau_3$, respectively}. The plotted values are inferred by \eqref{scaling} from the correlations between the optical Faraday rotation measurements. The Heisenberg uncertainty relation is represented by the value $1/2$. In (a) $\tau_3=2\text{ms}$, and in (b) $\tau_1=1\text{ms}$, and $\tau_4=1\text{ms}$ for both figures. The insets show the Wineland squeezing parameter $10\text{log}_{10}(\xi_R^2)$ in dB units for $\hat{x}_A^2$ and $\hat{p}_A^2$ (See Supplementary Materials). The error bars (one standard deviation) are derived from 5 identical experiments for $\Delta\hat{x}_A$ and $\Delta\hat{p}_A$ respectively, each consisting of 10000 repetitions of the corresponding pulse sequences.}
\end{figure*}

\noindent
\textbf{Experimental realization of retrodiction beyond HUR.}
The core of the experiment is a paraffin coated vapor cell~\cite{Balabas} containing about $10^{11}$ $^{87}$Rb atoms, as   sketched in Fig.~\ref{Fig:setup}(a). The anti-relaxation coating on the inner wall of the cell ensures a relatively long spin coherence lifetime. We initially populate the atoms in the state $|5S_{1/2},F=2, m_F =-2\rangle$ by optical pumping along the \emph{x}-direction parallel to the magnetic field \emph{B} with up to $97.9\%$ polarization, orienting the atoms along the \emph{x}-direction. It leads to a $6\%$ increase of the measured variance compared to the fully polarized CSS.  The projection noise limit is calibrated by measuring the noise of the collective spin of the unpolarized sample, which is 1.25 times that of the ideal CSS state (see Methods).

The linear-polarized probe light propagating along the \emph{z}-direction is turned on after the optical pumping, to measure quantum spin fluctuations in the transverse \emph{y}-\emph{z} plane through Faraday rotation. The intensity of the probe beam is stroboscopically modulated at twice the Larmor frequency in order to probe a spin component $\hat{x}_A(\theta)$\cite{Bao,GH}. The measurement direction $\theta$ is tuned by adjusting the phase of stroboscopic modulation. The phase is experimentally calibrated by radio frequency signal excitation (See Methods). As shown in Fig.~\ref{Fig:setup}(b), the probe pulse sequence is divided in four parts. The 1st pulse measures $\hat{p}_A$, representing the information of $\rho$. The 3rd and 4th pulse measure $\hat{x}_A$ and $\hat{p}_A$ respectively, representing the information of $E$. $\rho$ and $E$ together provide an estimation of the result at the time of 2nd pulse.  In each experimental repetition, we can only obtain the result for one value of $\theta$. Hence, the QND measurement of $\hat{x}_A(\theta)$ for different $\theta$ are obtained from independent experiments.

While Eq.\eqref{offdiag} permits evaluation of the conditional variance of the $m_2$ measurement, it does not take decoherence, losses and experimental imperfections into account. In the following, we shall present our bare experimental results based exclusively on analyses of correlations in the measurement data, cf. Eqs.(\ref{eq:corr1}, \ref{eq:corr2}) in Methods. These analyses only assume Gaussian correlations between the measurement outcomes, which is compatible with realistic errors and decoherence mechanisms in the system, and they assume that Eq.(\ref{scaling}) can be applied to infer the atomic variances from the optical measurement.


Fig.~\ref{Fig:xp} depicts the value of the uncertainty product $\Delta\hat{x}_A\cdot\Delta\hat{p}_A$ as a function of both measurement durations $\tau_1$ and $\tau_3$. The atomic variances are inferred from the variances of the optical measurements and the simple scaling \eqref{scaling} applicable for $\theta=0,\pi/2$. We find that the minimum value of $\Delta\hat{x}_A\cdot\Delta\hat{p}_A = 1/2\times(0.680\pm0.019)$ is smaller than the Heisenberg uncertainty limit of 1/2, and that it is obtained for $\tau_1=1$ ms and $\tau_3=2$ ms. Fig.~\ref{Fig:xp} shows that $\Delta\hat{x}_A\cdot\Delta\hat{p}_A$ first decreases with the probing time $\tau_1$ and corresponding measurement strength $\kappa_1$, while extending $\tau_1$ beyond $1$ ms causes incoherent scattering by spontaneous emission of the atoms, which reduces the effective mean spin projection, breaks the correlations between pairs of spins responsible for the squeezing, and adds random fluctuations to the collective ground state spin~\cite{KlausPRA2004,HammererPRA2004}. The later measurements of duration  $\tau_3$ and $\tau_4$ do not decohere the spin state at the earlier time $t$, but we observe a slow rise in the uncertainty product after an optimal probe duration $\tau_3$ in panel (b) of Fig.~\ref{Fig:xp}. The optimal probe duration is longer and the increase in uncertainty is slower than for $\tau_1$, which may be due to the reduction of the length of the mean spin, appearing in the definition of $\hat{x}_A$ and $\hat{p}_A$ and hence in the effective coupling strength of the field and atomic oscillator degrees of freedom. A too long $\tau_3$  depolarizes the spins and hence reduces the efficiency of the final $\tau_4$ measurement which plays an important role for the retrodicted variance (see Supplementary Material).

\begin{figure*}
\centering
\includegraphics[width=0.8\textwidth]{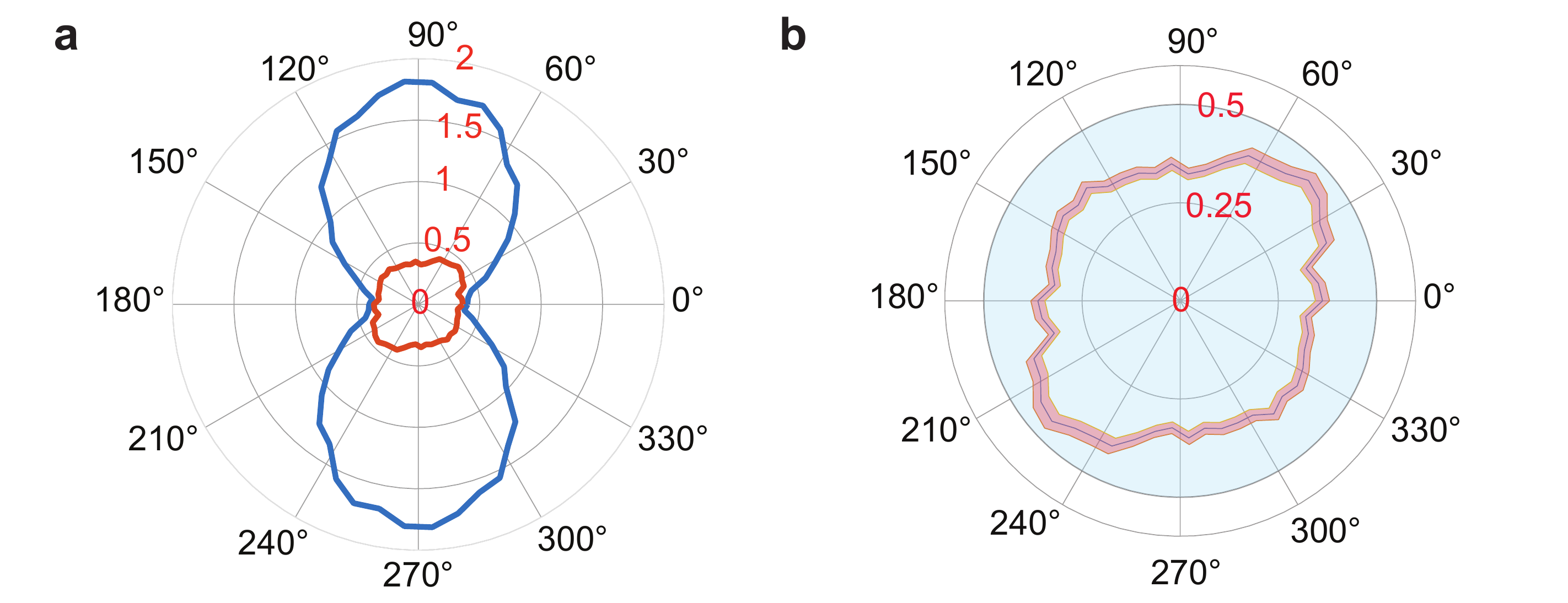}
\caption{\label{Fig:squeezing} \textbf{Polar plot of the experimentally observed (and rescaled) variances.} \textbf{a.} Measured $(\operatorname{Var}(m_2|m_1)-\frac{1}{2})/\kappa_2^2$  (blue) and $(\operatorname{Var}(m_2|m_1,m_3,m_4)-\frac{1}{2})/\kappa_2^2$ (red) for the Faraday probing of atomic quadratures  $\hat{x}_A(\theta)$.  The data are obtained with optical probe pulse durations corresponding to coupling strengths $\kappa_1^2=1.7$, $\kappa_2^2=0.81$, $\kappa_3^2=3.3$ and $\kappa_4^2=2.2$ between the atoms and the field. A value of $1/2$ corresponds to the standard quantum limit (SQL) {for the atomic spin variance}  \textbf{b.} Zoom-in of the red curve in (a). The shaded error bar (one standard deviation) is derived from 5 identical experiments, each consisting of 10000 repetitions of the pulse sequence.}
\end{figure*}

The outcome of the Faraday rotation measurements by coupling to atomic oscillator quadratures in arbitrary directions are shown by polar plots in Fig.\ref{Fig:squeezing}. The blue curve shows the variance conditioned on the prior measurements (Eq.\eqref{eq:corr1} in Methods). The result is scaled to atomic units, i.e., we plot $(\operatorname{Var}(m_2|m_1)-\frac{1}{2})/\kappa_2^2$, and we observe that the variance of the $\hat{p}_A$ quadrature (horizontal direction in the polar plot) is a factor $0.80\pm0.05$ below the SQL, and the anti-squeezed $\hat{x}_A$ quadrature is more noisy than the SQL (vertical direction in the polar plot). When we implement also the subsequent $m_3$ and $m_4$ measurements ((Eq.\eqref{eq:corr2} in Methods), the corresponding $(\operatorname{Var}(m_2|m_1,m_3,m_4)-\frac{1}{2})/\kappa_2^2$, shown as the red curve in Fig.~\ref{Fig:squeezing}(a)) shows that $\hat{x}_A$ is now subject to a similar squeezing effect as observed for  $\hat{p}_A$. The degree of squeezing is less than theoretically predicted for the ideal experiment in Fig.\ref{Fig:setup}(c) because decay and decoherence in the experiment is not considered in the simple theory.

Another reason for the discrepancy between Fig. \ref{Fig:setup}(c) and Fig. \ref{Fig:squeezing} is the more complex relations between the retrodicted variance of the optical measurement and the hypothetical projective atomic measurement for directions in the oscillator phase space, $\theta\neq 0,\frac{\pi}{2}$. The optical measurement outcome is correlated with the atomic observables, represented by the full matrix character of the operators $\rho$ and $E$ and hence its expectation value and variance in general involve both the probability distribution (diagonal elements) and coherences (off-diagonal elements) in Eq.\eqref{offdiag}. We thus emphasize that while the determination of the conditional variances of the optical measurement, for $\theta\neq 0,\frac{\pi}{2}$, are in agreement with our theoretical expectations for those quantities, the angular dependence of the results is different from the one of the hypothetical projective measurement of the atomic quadrature observables. Only in the limit of large $\kappa_2$, we recall that the optical Faraday rotation measurement becomes a projective atomic measurements, as shown in Supplementary Material.

\section*{Discussion}
We have demonstrated that measurements of two non-commuting observables of a spin oscillator can both be retrodicted with a precision below the SQL by prior and posterior QND detection of these observables. These measurements respectively condition the Gaussian Wigner fucntion for the density matrix $\rho(t)$ and the effect matrix $E(t)$ which together incorporate all our knowledge about the system at the past time $t$ and thus provide our best estimate of the outcome of any measurement on the system. The past quantum state theory violates the HUR for non-commuting observables of the spin oscillator, and it may also violate  error-disturbance relations \cite{Ozawa,BuschWerner} for sequential measurements of non-commuting observables, as the disturbance of the system by the first measurement does not prevent precise retrodiction of the second measurement outcome by the later probing. Our detection method is compatible with spin based sensing and spectroscopy \cite{SSRMP,Kasevich,James}, and the predicted and retrodicted evolution may offer insight and allow precision estimation of external influences, which can be applied to general quantum metrology \cite{Lloyd} such as interferometers \cite{Appel,Braverman} and magnetometers \cite{Budker-book,Romalis1,Bao} and force sensors based on mechanical oscillators \cite{MarkusRMP,Albert}.

\section*{\textbf{Acknowledgements}}
We thank M. Balabas and Precision Glassblowing (Boulder) for assistance in the vapor cell fabrication, and V. Vuleti\'c for helpful discussions. This work is supported by National Key Research Program of China under Grant No. 2016YFA0302000 and No. 2017YFA0304204, and NNSFC under Grant No. 61675047 and No. 91636107. K. M. acknowledges support from the Villum Foundation. H. S. acknowledges the financial support from the Royal Society Newton International Fellowship (NF170876) of United Kingdom.
\section*{\textbf{Competing interests}} The authors declare no competing interests.
\section*{\textbf{Additional information}}
\textbf{Correspondence and requests for materials} should be addressed to K.M., H.S. or Y.X.\\
\section*{\textbf{Data availability}}The datasets generated and analysed during this study are available
\clearpage
\section*{Methods}
\subsection*{Experiment details}
\textbf{Experimental setup.}
The experiment setup (Fig.~\ref{Fig:setup}) includes a 4-layer magnetic shielding, containing a paraffin-coated
$20~\text{mm}\times7~\text{mm}\times7~\text{mm}$ rectangular vapor cell, and a set of coils for generating a homogeneous bias magnetic field of $0.71~\text{G}$ which gives a ground-state Zeeman splitting of about $\Omega_L=2\pi\times500~\text{kHz}$. The measured decay time for the atomic Zeeman population and coherence are $T_1=125~\text{ms}$ and $T_2=20~\text{ms}$ respectively, with the latter mainly limited by residual magnetic field inhomogeneity. A $y$-polarized probe laser propagating along the $z$ axis is blue-detuned by $2.1~\text{GHz}$ from the $5S_{1/2},F=2\rightarrow 5P_{3/2},F'=3$ transition of the D2 line. Its intensity is modulated at twice the Larmor frequency by an acousto-optic-modulator to implement the stroboscopic quantum back-action evasion protocol~\cite{GH}, with an optimal duty cycle of $14\%$.  In this protocol, the variance of $\hat{S}_{y}$ in photon shot noise unit for pulse duration $\tau$ after the interaction is \cite{GH}
\begin{equation}
\begin{aligned}\label{strobnoise}
\text{Var}(\hat{S}_{y,\tau}^{out})_{SN}&\approx \big[ 1+\tilde{\kappa}^2+\frac{\tilde{\kappa}^4}{3}\frac{1-\text{Sinc}(\pi D)}{1+\text{Sinc}(\pi D)} \big],
\end{aligned}
\end{equation}
where $D$ is duty cycle, and $\tilde{\kappa}^2$ is proportional to ${\kappa}^2$ with a coefficient accounting for the stroboscopic effect (See Supplementary Materials). The home-made balanced photo detector for measuring the $S_y$ has a quantum efficiency of $92.4\%$ and it operates in the unsaturated regime up to 12 mW.

First, we prepare the atoms in the state $5S_{1/2}\left | F=2, m_F =-2\right \rangle$ (with quantum number $m_F$ associated with the quantization axis along $x$, the direction of the magnetic field) by applying the circular polarized and spatially-overlapped $\sigma ^{-}$ pump and repump lasers propagating along the \emph{x}-direction. We achieve up to $97.9\%$ degree of spin orientation, as measured by the magneto-optical resonances~\cite{MORS}. The optimized laser powers are 50~mW for the repump and 5~mW for the pump, both having elongated-Gaussian transverse intensity distribution. The probe mode is a symmetric Gaussian with $1/e^2$ beam diameter of 6 mm. All three fields cover nearly the entire cell volume.

\noindent
\textbf{Calibration of the spin projection noise limit.}
The coupling strength $\tilde{\kappa}^2$ of the atom and field variables is calibrated by measuring the spin noise of the atomic ensemble in thermal equilibrium which is unpolarized and not affected by the tensor interaction. The observed spin noise in the thermal state should be $\frac{5}{4}\cdot \text{Var}(\hat{x}_{A},\hat{p}_{A})_{CSS}$ by the following reasoning. The thermal state is isotropic, which implies $\langle \widehat{j}_x^2\rangle=\langle\widehat{j}_y^2\rangle=\langle\widehat{j}_z^2\rangle=\frac{F(F+1)}{3}=2$ for $F=2$. Meanwhile, all sublevels have the same population, including those in $F=1$ which are not observed in the measurement. Since there are 8 sublevels in total for $5S_{1/2}$ hyperfine states, and 5 of them belong to $F=2$, the observed noise will thus be $2\cdot \frac{5}{8}=\frac{5}{4}$.
This is $\frac{5}{4}$ times the noise variance $\langle\widehat{j}_z^2\rangle$ in the CSS deduced from the Heisenberg uncertainty relation.

In our system, the linearly polarized probe light is transmitted through the atomic sample and it undergoes a small polarization rotation due to its interaction with the atoms. Our effective polarization homodyne detection employs the original linear polarization component as a local oscillator for the orthogonal component of the field system generated by the interaction with the atoms. The total noise on the optical readout signal includes photon shot noise and spin noise, $\sigma^2_{x_{ph}} = \frac{1}{2} + \kappa^2 \sigma^2_{p_{at}}$. In addition to estimating the coupling strength $\kappa$ from the physical parameters, we can thus infer its value from the noise in the probe experiments. In order to calibrate the photon shot noise level, the Larmor frequency is tuned far away from the lock-in detection bandwidth by changing the DC bias magnetic field, suppressing the noise contribution from the spin oscillator. The photon shot noise depends linearly on the input probe power \cite{Bao}, since for the coherent state of light the variances of $\hat{S}_y$ and $\hat{S}_z$ should satisfy  $\text{Var}(\hat{S}_y)=\text{Var}(\hat{S}_z)=\frac{S_x}{2}$. In addition, as shown in Ref. \cite{Bao}, the linear scaling of spin noise power as a function of atomic number indicates a quantum-limited performance and the QND character of the measurement.

\noindent
\textbf{Calibration of the measurement direction in the spin oscillator phase space.}
In the rotating frame, we first measure the quadrature $\hat{p}_A$. To measure $\hat{x}_A$, we need to wait for the spin oscillator to rotate by $\pi/2$, i.e., a quarter of the Larmor period. Here we describe how we verify that the measured quadrature is really $\hat{x}_A$, perpendicular to $\hat{p}_A$.

As depicted in the main text, we use a homogeneous DC bias magnetic field $B_x$ in the spin orientation direction, which is the \emph{x} direction in this paper. This corresponds to an additional Hamiltonian term $\hat{H}=\Omega_L\hat{J}_x$ with $\Omega_L=g_F\mu_B B_x/\hbar$, where $g_F$ is the hyperfine Land$\acute{e}$ g-factors for the ground state of $^{87}$Rb, while $\mu_B$ and $B$ are the Bohr magneton and the magnitude of the applied magnetic field. If we also add a radio frequency (RF) magnetic field oscillating at frequency $\Omega$ along the \emph{y} direction such that in the absence of light
$\textbf{B}_{ext}=B_x\textbf{e}_x+[B_c\cos (\Omega t+\phi)+B_s\sin (\Omega t+\phi)]\textbf{e}_y$ with constant $B_c$ and $B_s$ one can derive Heisenberg equations of motion for the collective spin components $\hat{J}'_y$ and $\hat{J}'_z$ in the rotating frame \cite{HS_thesis},
\begin{equation}
\begin{aligned}
\frac{\partial \hat{J}'_y }{\partial t}&=-\omega_s\sin (\Omega_L t)\sin (\Omega t+\phi)J_x\\
\frac{\partial \hat{J}'_z }{\partial t}&=-\omega_c\cos (\Omega_L t)\cos (\Omega t+\phi)J_x,
\end{aligned}
\end{equation}
with $\omega_{c,s}=g_F\mu_B B_{c,s}/\hbar$. Choosing the phase and the frequency of the RF-drive such that $\phi=0$ and $\Omega=\Omega_L$, we obtain $\frac{\partial \hat{J}'_y }{\partial t}=-\frac{\omega_sJ_x}{2}$ and $\frac{\partial \hat{J}'_z }{\partial t}=-\frac{\omega_cJ_x}{2}$, given interaction durations $T$ satisfying the condition $\omega_{c,s}T\ll 1\ll \Omega T$. With the RF magnetic field pulses we are thus able to independently change the spin components $\hat{J}'_y$ and $\hat{J}'_z$ by an amount controlled by the sine and cosine components $B_s$ and $B_c$.

Therefore, we can implement the calibration with the pulse sequence depicted in Extended Data Fig. 1. An RF field pulse is applied between the pump laser and probe laser to create a transverse spin excitation, rotating in the $y$-$z$ plane. The stroboscopically modulated probe laser is separated into two parts. The first measures the projection component in $\hat{p}_A$ and the second measures that in $\hat{x}_A$. The phase of the RF field determines the phase of the rotating transverse spin (i.e. the direction in the rotating frame). It is easy to maximize the signal of the first part of probe by adjusting the phase of the RF field. At this time, the induced transverse spin should be in the $\hat{p}_A$ direction (in the rotating frame) and there should be only a minimal spin component in the $\hat{x}_A$ direction. We verify that, after a quarter of the Larmor period, the induced spin is in the $\hat{x}_A$-direction.

\noindent
\textbf{Data analysis.}
Since the expectation value $\mu$ of the gaussian distribution $Pr(m_2|m_1)$ is proportional to $m_1$, experimentally, the variance of $m_2$ conditioned on the measurement before $t$, with result $m_1$, should be obtained from linear numerical feedback \cite{RMP} as
\begin{equation}
\begin{aligned}
&\text{Var}(m_2|m_1)=min_{\alpha}\left[\text{Var}(m_2-\alpha m_1)\right]\\
&=min_{\alpha}[\text{Var}(m_2)+\alpha^2\text{Var}(m_1)-2\alpha\text{Cov}(m_2,m_1)]\\
&=\text{Var}(m_2)-\frac{\text{Cov}^2(m_2,m_1)}{\text{Var}(m_1)}.
\end{aligned}
\end{equation}
where the minimum is achieved when
\begin{equation}
\begin{aligned}
\alpha=\frac{\text{Cov}(m_2,m_1)}{\text{Var}(m_1)}
\end{aligned}
\end{equation}
Similarly, since the expectation value $\mu_{\rho E}$ of the gaussian distribution $Pr(m_2|m_1,m_3,m_4)$ is proportional to $m_1$, $m_3$ and $m_4$, the variance of $m_2$ conditional on the measurement before and after $t$, i.e. $m_1$, $m_3$ and $m_4$, should also be obtained from linear numerical feedback as
\begin{equation}\label{eq:corr1}
\begin{aligned}
&\text{Var}(m_2|m_1,m_3,m_4)=min_{\alpha,\beta,\gamma}[\text{Var}(m_2-\alpha m_1-\beta m_3-\gamma m_4)]\\
&=min_{\alpha,\beta,\gamma}[\text{Var}(m_2)+\alpha^2\text{Var}(m_1)+\beta^2\text{Var}(m_3)+\gamma^2\text{Var}(m_4)\\
&-2\alpha\text{Cov}(m_2,m_1)-2\beta\text{Cov}(m_2,m_3)-2\gamma\text{Cov}(m_2,m_4)\\
&+2\alpha\beta\text{Cov}(m_1,m_3)+2\alpha\gamma\text{Cov}(m_1,m_4)+2\beta\gamma\text{Cov}(m_3,m_4)]\\
\end{aligned}
\end{equation}
The minimal is achieved when
\begin{equation}\label{eq:corr2}
\begin{aligned}
\alpha&=\frac{1}{\Lambda}(\text{Cov}_{14}\text{Cov}_{23}\text{Cov}_{34}+\text{Cov}_{13}\text{Cov}_{24}\text{Cov}_{34}-\text{Cov}_{12}\text{Cov}_{34}^2\\
&-\text{Cov}_{14}\text{Cov}_{24}\text{Cov}_{33}-\text{Cov}_{13}\text{Cov}_{23}\text{Cov}_{44}+\text{Cov}_{12}\text{Cov}_{33}\text{Cov}_{44})\\
\beta&=\frac{1}{\Lambda}(\text{Cov}_{13}\text{Cov}_{14}\text{Cov}_{24}+\text{Cov}_{12}\text{Cov}_{14}\text{Cov}_{34}-\text{Cov}_{23}\text{Cov}_{14}^2\\
&-\text{Cov}_{24}\text{Cov}_{34}\text{Cov}_{11}-\text{Cov}_{12}\text{Cov}_{13}\text{Cov}_{44}+\text{Cov}_{11}\text{Cov}_{23}\text{Cov}_{44})\\
\gamma&=\frac{1}{\Lambda}(\text{Cov}_{13}\text{Cov}_{14}\text{Cov}_{23}+\text{Cov}_{12}\text{Cov}_{13}\text{Cov}_{34}-\text{Cov}_{24}\text{Cov}_{13}^2\\
&-\text{Cov}_{23}\text{Cov}_{34}\text{Cov}_{11}-\text{Cov}_{12}\text{Cov}_{14}\text{Cov}_{33}+\text{Cov}_{24}\text{Cov}_{11}\text{Cov}_{33})\\
\Lambda&=2\text{Cov}_{13}\text{Cov}_{14}\text{Cov}_{34}-\text{Cov}_{14}^2\text{Cov}_{33}-\text{Cov}_{13}^2\text{Cov}_{44}\\
&+\text{Cov}_{11}\text{Cov}_{33}\text{Cov}_{44}
\end{aligned}
\end{equation}
Here $\text{Cov}_{uv}=\text{Cov}(m_{u},m_{v})$ represent the covariance between $m_{u}$ and $m_{v}$, where $u,v=1,2,3$. When $u=v$, $\text{Cov}_{uu}=\text{Var}(m_{u})$.
\begin{figure*}
\centering\label{Exd1}
\includegraphics[width=0.8\textwidth]{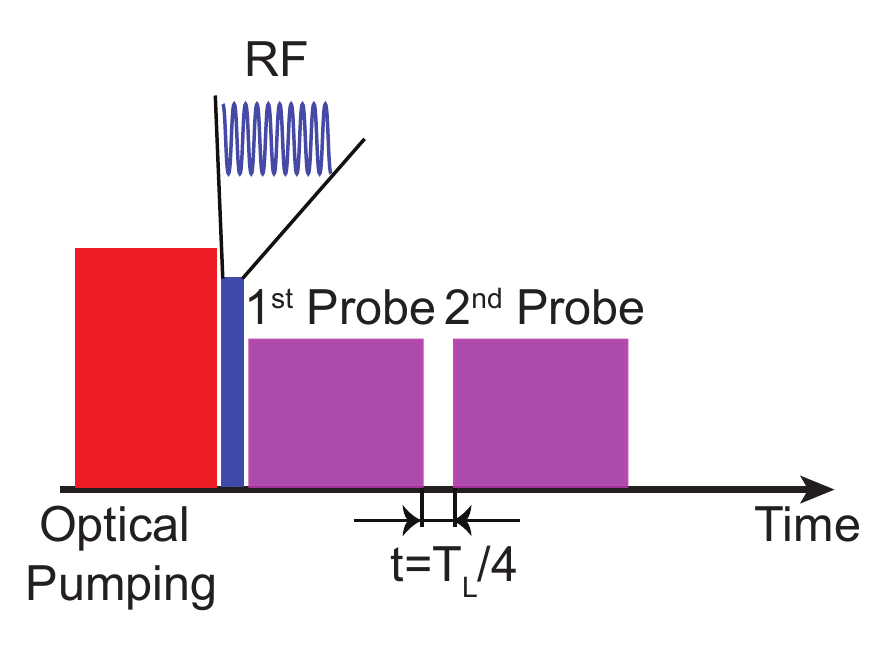}\par
\textbf{Extended DataFig.1. Pulse sequence for calibration of measurement direction.}
\end{figure*}

\end{document}